\documentclass[sigconf]{acmart}
\usepackage{enumerate}
\usepackage{tabularx}
\usepackage{multirow}
\usepackage{float}
\usepackage{graphicx} 
\usepackage{enumitem}
\usepackage[normalem]{ulem}
\usepackage{subfig}

\AtBeginDocument{%
  \providecommand\BibTeX{{%
    \normalfont B\kern-0.5em{\scshape i\kern-0.25em b}\kern-0.8em\TeX}}}

\begin{document}

\title{Two Truths and a Lie: Exploring Soft Moderation of COVID-19 Misinformation with Amazon Alexa}


\author{Donald L. Gover}
\affiliation{%
  \institution{DePaul University}
  \streetaddress{ 1 E Jackson Blvd}
  \city{Chicago}
  \state{Illinois}
  \country{USA}}
\email{dgover@mail.depaul.edu}

\author{Filipo Sharevski}
\affiliation{%
  \institution{DePaul University}
  \streetaddress{ 1 E Jackson Blvd}
  \city{Chicago}
  \state{Illinois}
  \country{USA}}
\email{fsharevs@cdm.depaul.edu}

\renewcommand{\shortauthors}{Authors}

\begin{abstract}
In this paper, we analyzed the perceived accuracy of COVID-19 vaccine Tweets when they were spoken back by a third-party Amazon Alexa skill. We mimicked the soft moderation that Twitter applies to COVID-19 misinformation content in both forms of warning covers and warning tags to investigate whether the third-party skill could affect how and when users heed these warnings. The results from a 304-participant study suggest that the spoken back warning covers may not work as intended, even when converted from text to speech. We controlled for COVID-19 vaccination hesitancy and political leanings and found that the vaccination hesitant Alexa users ignored any type of warning as long as the Tweets align with their personal beliefs. The politically independent users trusted Alexa less then their politically-laden counterparts and that helped them accurately perceiving truthful COVID-19 information. We discuss soft moderation adaptations for voice assistants to achieve the intended effect of curbing COVID-19 misinformation.
\end{abstract}



\maketitle

\section{Introduction}
Misinformation and rumors about viruses are not a new thing. What's new is the size of the target audience and the speed with which such damaging information spreads online. It took weeks and a publication in foreign media outlets for the KGB-initiated ``Operation Infektion,'' spreading the rumour that HIV/AIDS was a misfired American biological weapon in the 1980s, to gain traction and manipulate the public opinion, at least outside of the U.S. \cite{Boghardt}. Fast forward 40 years, one can choose an aspect of the novel COVID-19 virus misinformation (e.g. origin, vaccines' adverse effects, efficacy) as well as audience (e.g. Twitter, Facebook, Parler) and manipulate the public opinion in a matter of seconds both in the U.S. and everywhere in the world \cite{Jachim, Pieroni}.   

While there was no structure of countering the old-style vaccination rumors, a coordinated soft moderation of misinformation was rather quickly put in place during the COVID-19 pandemic on the mainstream social media platforms \cite{Roth}. Such an effort was warranted because the real world implications of (mis)information and unverified rumors have a direct impact on public health, especially of mass immunization. According to Twitter, who opted to use warning tags and covers as misinformation labels, the supposed aim of the soft moderation is to reduce misleading or harmful information that could ``incite people to action and cause widespread panic, social unrest or large-scale disorder'' \cite{Roth}. 

Consequently, the academic community set forth to investigate whether this seemingly noble aim was in fact achieved among the social media users and the general public. In an early study, the soft moderation labels were found to have a counter effect since the warnings convinced people to believe the misinformation even more than if the labels were not there \cite{Clayton}. A study exploring this so-called ``soft moderation'' implemented by Twitter found that the platforms' content with warning labels generated more action than content without said labels \cite{Zannettou}. The soft moderated Tweets were more likely to be distributed than a ``valid'' Tweet through discourse, not always because the soft moderated Tweets contained misinformation but because they contained a response to mock or disclaim an original author or a valid Tweet. Interestingly, a mere 1\% of the Tweets gathered for the study were labeled with a COVID-19 warning and a number of these few Tweets were found to be mislabeled simply because they contained the words ``oxygen'' and ``frequency''. Another study, in this context, found that some users did not trust the soft moderation intervention and felt that Twitter itself was biased and mislabeling content \cite {Geeng}. Even without the warning labels, a varying degree of users’ perceptions and beliefs regarding vaccines in general and about COVID-19 vaccines in particular plays a role in individuals' reaction to (mis)information Tweets. A study on vaccine misinformation spreading and acceptance on social media platforms showed the effectiveness of ``influential users'' within like minded vaccine-fearing followers to be very high \cite{Featherstone20}. 

Mainstream social media platforms like Twitter allow for visual discernment of the information including the warning labels, formation of so-called ``influencer'' accounts, and direct communication of the engagement with the content metrics such as number of replies, re-Tweets, likes, and shares. While all of these factors certainly affect the receptivity of any COVID-19 vaccine information, little attention is devoted to exploring how people respond to soft moderated COVID-19 vaccine Tweets when these are delivered through a voice assistant like Amazon Alexa. Unlike the traditional Twitter interface, Alexa is the sole authority or ``influencer'' that delivers the Tweet when prompted possibly without disclosing the details of any labeling or engagement metrics (depending on the configuration for retrieval and presentation of one's twitter diet). Users usually trust Alexa and worry only about Alexa intruding into their privacy, but not about the validity of the information delivered by Alexa \cite{Lau}. Since the configuration of how one's twitter diet is spoken-back by Alexa can be customized and implemented by various third-party applications, called ``skills'' we set to explore how users will respond when such a third-party skill is used to deliver soft moderated COVID-19 vaccine Twitter content. Studies in the past have shown that third-party skills could be dangerous in that can silently rephrase information from any source to mislead a user and induce misperception about a polarizing topic such as vaccination, free speech, or government actions  \cite{mimTweet, mimfacebook, Malexa}.

\section{Soft Moderation by Alexa}
Malicious skills for Alexa, unsurprisingly, exist on the Amazon Skills Store \cite{Vaidya}. Some of them pray on malicious voice commands (songs, phonemes, inaudible noises) to be interpreted as legitimate by Alexa and invoke a skill that snoops on the user or attempts to steal their credentials \cite{Carlini, Kumar, Xuejing}. Some of them try to lure the user to spell their password directly to Alexa in the voice variant of phishing or ``vishing'' \cite{srl}. Some of them don't bother with the invocation logic or users' credentials, but instead, manipulate the content that is spoken back to the users \cite{Sharevski}. This new type of malicious skills, in a man-in-the-middle fashion, rewords a legitimate content in such way that the user makes an improbable interpretation of a set of true facts. The goal of this skill is to induce misperception, that is, to distort the reality of a targeted user in a similar way information operations attempt to do with disseminating alternative narratives on social media \cite{Starbird}. The authors of this proof-of-concept  malicious skill, manipulated regulatory news to present the government in two different lights as a pro-worker and as a pro-business. A user study was conducted that validated the potential of the malicious skill to induce misperception about the government's proclivity towards workers and the businesses.  

The possibility for manipulating spoken back content by the voice assistant prompted us to think how such a skill could be weaponized if the content comes from Twitter. Twitter is one the main ``battlefileds'' for information operations with numerous alternative narratives surrounding the COVID-19 pandemic. Recognized that these alternative dangerously pollute the public discourse on the platform and started to actively label them as ``COVID-19 misinformation'' \cite{Roth}. The labels come in two forms: (1) a warning cover with a verbose text hiding the text and allowing the user to skip a COVID-19 misinformation Tweet; and (2) a warning tag underneath a content providing a link for users to ``get the facts about COVID-19.'' Twitter is not just providing content but also provides context on it's truthfulness to the user with such label. It has been found that similar labels on social media reduce the perceived accuracy of an alternative narrative, e.g. it help users recognize an attempt for inducing misperception \cite{Clayton}. 

Circling back to the possibility for inducing misperception by manipulating content by a malicious skill, we are also interested in the ability of this skill to suppress or deliberately insert such labels in order to enhance or attenuate the misperception-inducing effect of a selected Twitter content. In other words, we wanted to know how the user will perceive a spoken back Twitter content by the malicious skill in the following scenarios: (1) Alexa utters the warning label text and speaks back the content of a misinformation/valid Tweet; (2) Alexa speaks back the content of a misinformation/valid Tweet and then utters the warning tag; and (3) Alexa suppresses the warning label text/tag and only speaks back the content of a misinformation/valid Tweet;

\section{Alexa Tweeter Reader Skill}
\begin{figure*}
  \includegraphics[width=\textwidth,height=5cm]{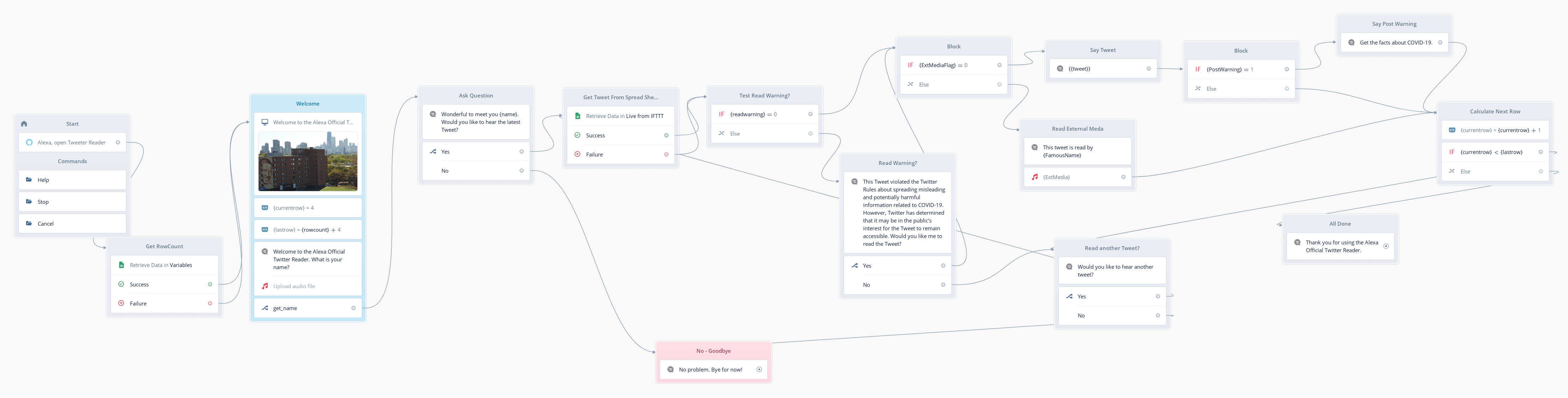}
  \caption{The Alexa Twitter Reader Skill (logical flow)}
    \label{fig:voiceflowapp}
\end{figure*}

The thrust of our research was not to fully implement a malicious Alexa skill that reads Tweets but rather to test the effect should one be developed (we elaborate the ethics of our research in the Discussion section), therefore, we produced a proof-of-concept that closely resembled such a skill. The ``Twitter Reader'' skill could automatically interact with both the user and Twitter adjusting the interaction depending on the ``blueprint'' selected for the user interaction \cite{alexa_skill_blueprints}. The Amazon Alexa skill ecosystem has a broad selection of ``blueprints'' or prepackaged skills from which one can easily deploy voice activated applications to the Alexa with minimal effort. Initially for this study we started using the Flash Briefing skill which allowed us to present to a user a predefined set of Tweets. This however was limited as having to statically define the Tweets. The requirements for dynamic content retrieved from Twitter, possible presentation of multimedia files, and conditional responses quickly outgrew the Flash Briefing skill and a more sophisticated platform was required. We migrated our implementation to VoiceFlow where we developed the Twitter Reader skill flow shown in Figure \ref{fig:voiceflowapp} \cite{voiceflow}.

To invoke the Twitter Reader skill the user utters the skill’s invocation name \texttt{Alexa, open the Twitter Reader}. Alexa response with the welcome message \texttt{Welcome to the Alexa Official Twitter Reader} and ask the name of the user. This is done to allow for reading Tweets from multiple accounts in one household, for example. Once Alexa connects the name to the configured Twitter account, Alexa prompts the user whether \texttt{they want to hear the latest Tweet in their Tweeter stream}. Assuming a confirmation answer from the user, Alexa retrieves the Tweet together with any warning labels attached to it. If a COVID-19 misinformation warning label is appended at the end of the Tweet, Alexa first converts the Tweet's text into speech, clearing special characters like ``@'' for account links and ``\#'' for hashtags, and responds, for example \texttt{A second nytimes article quotes doctors who say the mRNA technology used in COVID vaccines may cause immune thrombocytopenia, a blood disorder that last month led to the death of a Florida doctor after getting the Pfizer vaccine} (this Tweet is also shown in Figure 2a). Once done reading the Tweet, Alexa appends the warning tag, removing the exclamation mark favicon, and utters \texttt{Get the information about COVID nineteen}. 

If a warning cover precedes the Tweet, Alexa first converts the text of the warning to speech and utters it back to the user. In the case of COVID-19 misinformation, this cover reads: \texttt{This Tweet violated the Twitter Rules about spreading misleading and potentially harmful information related to COVID-19. However, Twitter has determined that it may be in the public's interest for the Tweet to remain accessible. Would you like me to read the Tweet?}. If the user says \texttt{Yes}, Alexa proceeds and converts the Tweet's text into speech as in the scenario shown in Figure 3. If the user says \texttt{No}, then Alexa responds \texttt{Would you like to hear another Tweet?}. If the user again declines to hear another Tweet, Alexa simply responds \texttt{Thank you for using the Alexa Official Twitter Reader}. Otherwise, Alexa proceeds and selects the next Tweet in the stream.


\section{Research Study}
\subsection{Misperceptions: Accuracy of Tweets}
To test the misperception inducing-effect we investigated whether (mis)information Tweets about the COVID-19 vaccine efficacy, spoken back by Alexa and in the presence or absence of warning labels affect individuals’ perceptions of the Tweet’s accuracy with the following set of hypotheses: 

\begin{itemize}[leftmargin=*]
\itemsep 1em
    \item H1: The utterance of a warning tag following a Tweet containing \textit{misleading} information about COVID-19 vaccines by Alexa will not reduce the perceived accuracy of the spoken back Tweet's content relative to a no warning tag condition. 

    \item H2: The utterance of a warning cover before a Tweet containing \textit{misleading} information about COVID-19 vaccines by Alexa will not reduce the perceived accuracy of the spoken back Tweet's content relative to a no warning cover condition.
        
    \item H3: The utterance of a warning tag following a Tweet containing \textit{valid} information about COVID-19 vaccines by Alexa will not reduce the perceived accuracy of the spoken back Tweet's content relative to a no warning tag condition. 

    \item H4: The utterance of a warning cover before a Tweet containing \textit{valid} information about COVID-19 vaccines by Alexa will not reduce the perceived accuracy of the spoken back Tweet's content relative to a no warning cover condition.
\end{itemize}

To test the first hypothesis we utilized the Tweets containing \textit{misleading information} shown in Figure 2a and Figure 2b. The Tweet in Figure 2a has a warning tag indicating that the above content is misinformation. The Tweet promulgates COVID-19 rumour about a rare adverse effect that was linked to the SARS-CoV-2 virus, not the vaccine, at the time of writing \cite{Chappell}. To remove bias due to the ``influencer'' effect, the Tweet comes from a verified account ``TheVaccinator'' (which we made up). An alteration of the same Tweet is shown in Figure 2b without the accompanying warning tag. To test the second hypothesis we utilized the Tweets shown in Figure 2b and Figure 3 (which includes a warning cover instead of a warning tag). In all the cases, the content of the Tweet and the warning covers/tags were converted from text-to-speech by the malicious Alexa skill as described above.  

\begin{figure}[htbp]
  \centering
  \includegraphics[width=\linewidth]{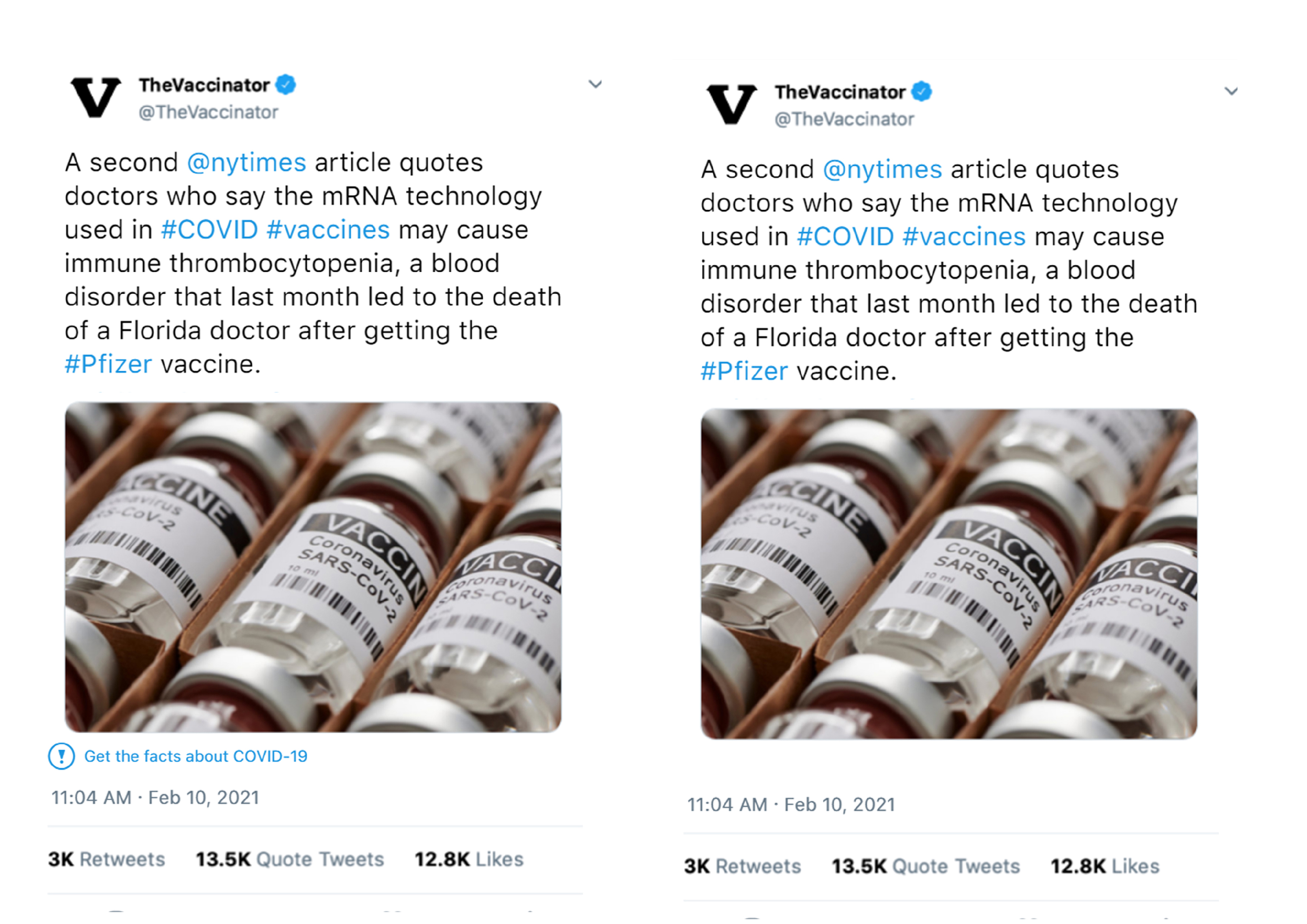}
  \subfloat[\label{MI with tag}]{\hspace{.5\linewidth}}
  \subfloat[\label{MI without tag}]{\hspace{0.5\linewidth}}
  \caption[MI Tweet]{A Misleading Tweet: (a) \textit{With} a Warning Tag; \\(b) \textit{Without} a Warning Tag for Misleading Information.\label{MI Tweet}}
\end{figure}

\begin{figure}[htbp]
  \centering
  \includegraphics[width=\linewidth]{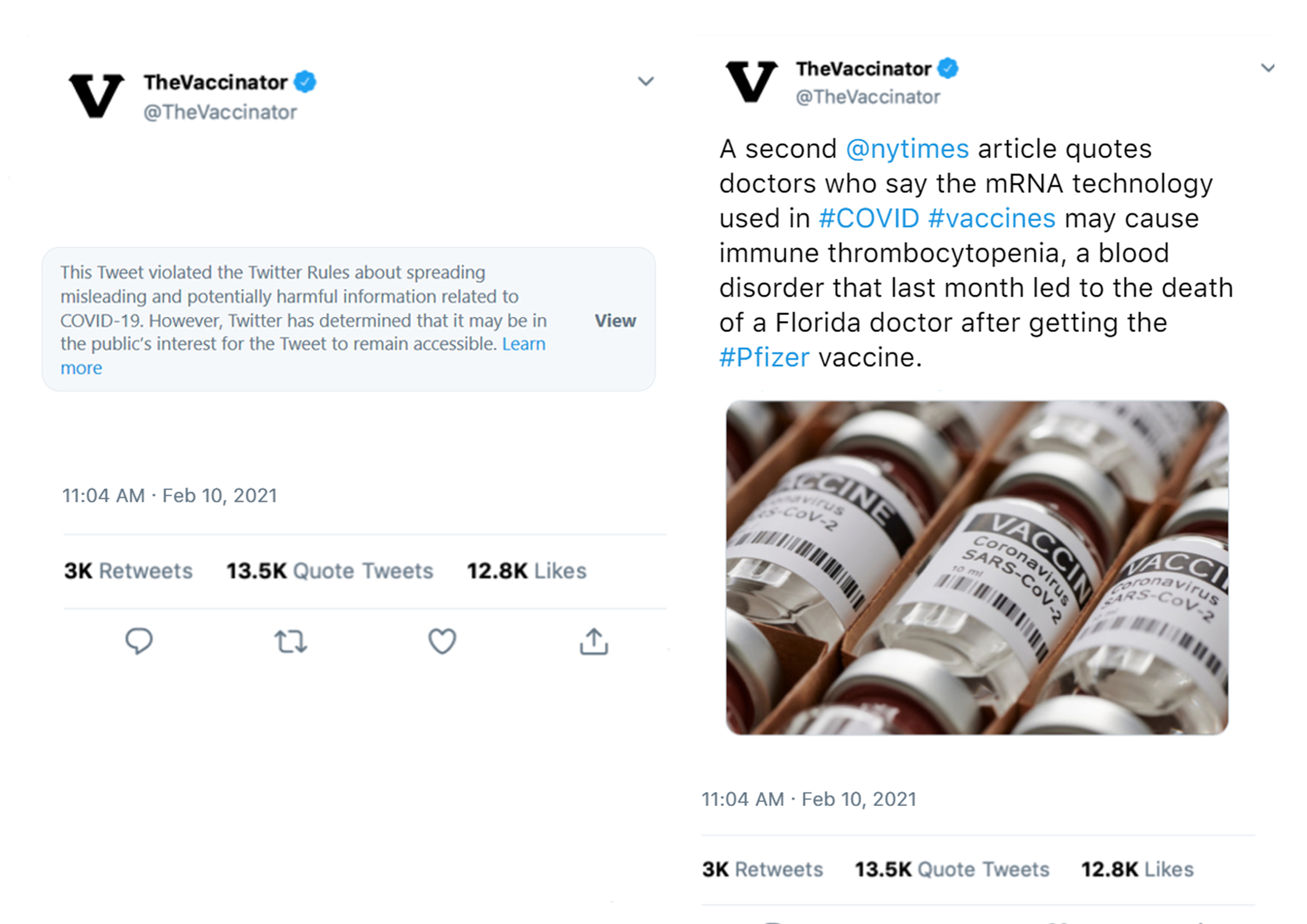}
  \caption{A Warning Cover Preceding a Misleading Tweet}
\end{figure}

To test the third hypothesis we utilized the Tweets containing \textit{verified information} shown in Figure 4a and Figure 4b. The original Tweet content cites the CDC's announcement about proceeding with the second dose of the COVID-19 vaccine in case an individual has a serious reaction from the first dose, altered to include a warning tag in Figure 4b \cite{CDC}. To control for bias, the Tweet comes from a verified account ``TheVirusMonitor'' instead of the CDC account. To test the fourth hypothesis we utilized the Tweets containing \textit{verified information} shown in Figure 4a and Figure 5. We retained Figure 4a for the comparison of the conditions and altered the labeling in the Figure 4b to include a warning cover instead of a warning tag. Similarly, the text from the twitter content and the labels was converted as a text-to-speech by Alexa.

\begin{figure}[htbp]
  \centering
  \includegraphics[width=\linewidth]{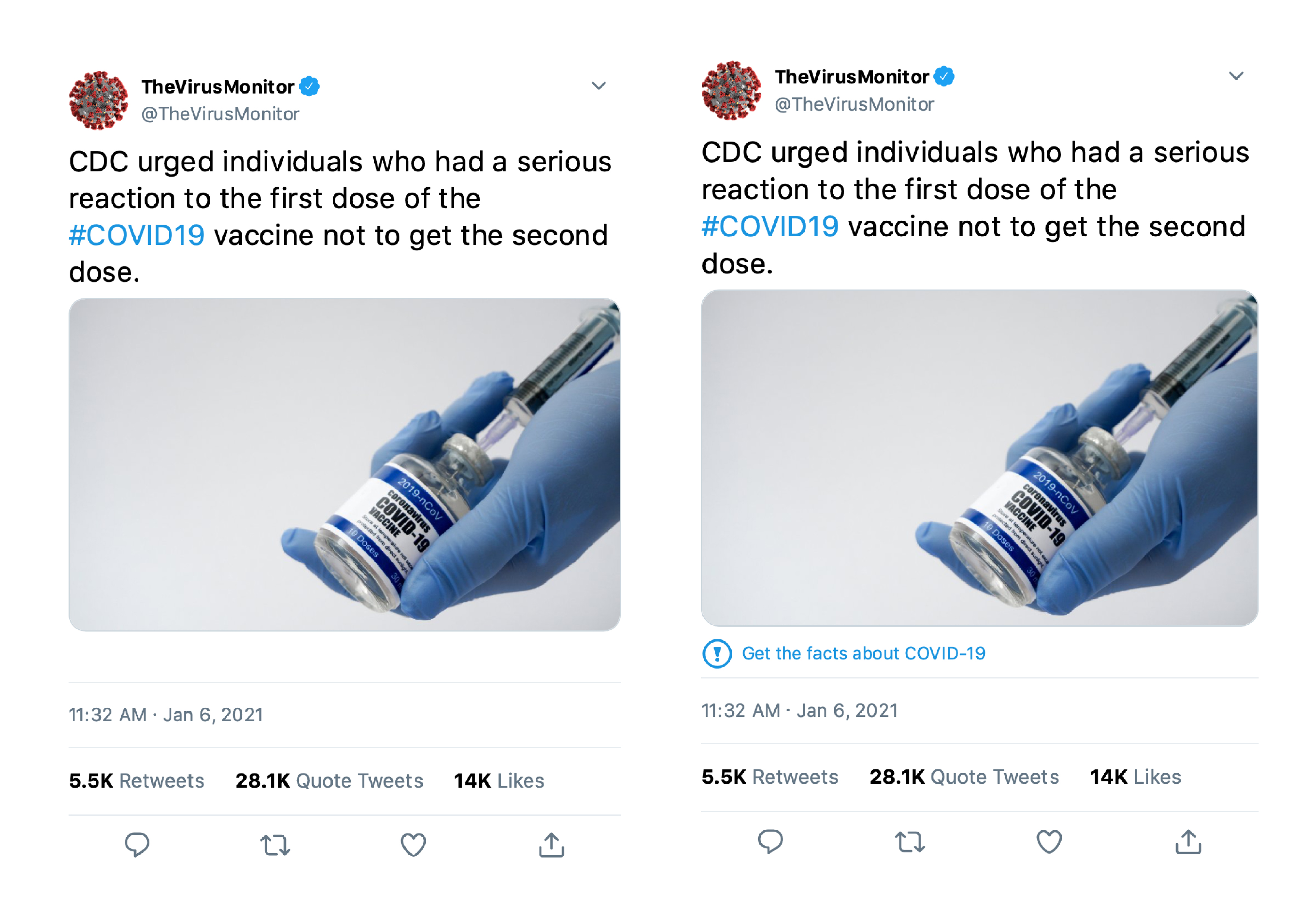}
  \subfloat[\label{VI without tag}]{\hspace{.5\linewidth}}
  \subfloat[\label{VI with tag}]{\hspace{.5\linewidth}}
  \caption[VI Tweet]{A Verified Tweet: (a) \textit{Without} a Warning Tag; \\ (b) \textit{With} a Warning Tag for Misleading Information.\label{VI Tweet}}
\end{figure}

\begin{figure}[h]
  \centering
  \includegraphics[width=\linewidth]{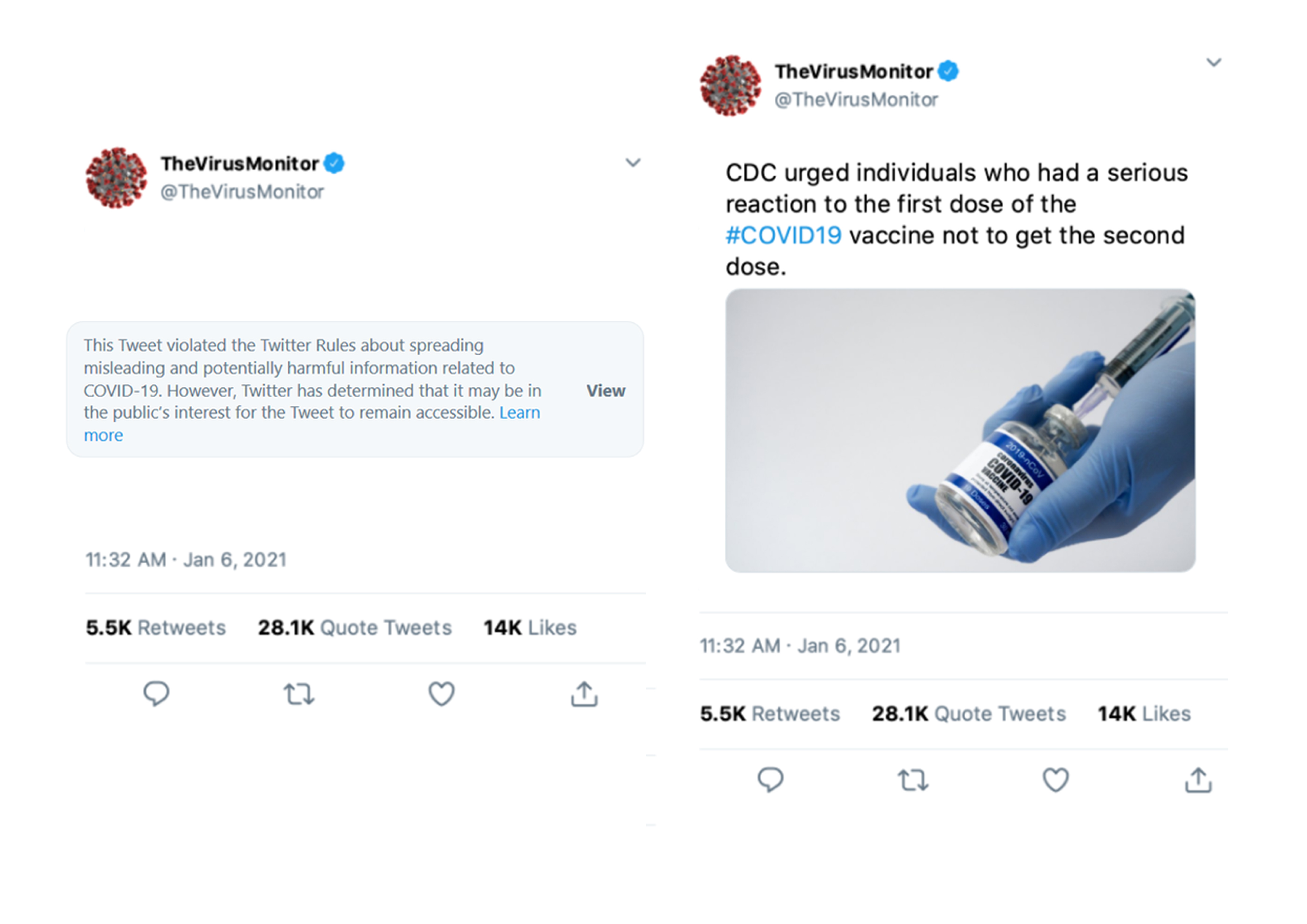}
  \caption{A Warning Cover Preceding a Verified Tweet}
\end{figure}

\subsection{Misperceptions: Hesitancy and Political Leanings}
We also controlled for hesitancy and political leanings, following the moderating effect of these variables reported in \cite{Zannettou}, \cite{Pennycook1}, to examine the perceived accuracy of spoken-back Tweets with COVID-19 vaccine information in presence/absence of uttering warning labels. We used the same Tweets as shown in Figures 2-5 together to test the following hypotheses:  

\begin{itemize}[leftmargin=*]
\itemsep 1em

        \item H5a: The COVID-19 vaccine personal hesitancy will not affect the perception of accuracy of a spoken back Tweet with \textit{misleading} information about COVID-19 in any condition (with a warning a preceding warning cover, with a following warning tag, or without any warning)

        \item H5b: The COVID-19 vaccine personal hesitancy will not affect the perception of accuracy of a spoken back Tweet with \textit{valid} information about COVID-19 in any condition (with a warning a preceding warning cover, with a following warning tag, or without any warning)
        
\end{itemize}

To test the association between one's political leanings and the perceived accuracy of the Tweets from Figures 2-5 we asked: 

\begin{itemize}[leftmargin=*]
\itemsep 1em
        \item RQ1: Is there a difference in the perceived accuracy of spoken-back COVID-19 misleading/verified Tweets with warning labels (tags or covers) between Republican, Democrat, and Independent users?
        
\end{itemize}

\subsection{Sampling and Instrumentation}
We set to sample a population using Amazon Mechanical Turk that is 18 years or above old, is a Twitter user and an Amazon Alexa user, and has encountered at least one Tweet into their feed that relates to COVID-19 vaccines. Because we were not allowed to physically invite the participants in our lab, we recorded the interaction between a user prompting Alexa to open the Twitter Reader skill and read the Tweets, which was offered as a recording to each participant. The study was approved by the Institutional Review Board (IRB) as a non-full disclosure experiment. Consequently, participants were initially told that they are being asked to gauge the effectiveness or usability of the Alexa skill as an experimental COVID-19 Tweeter Reader. After participation, each participant was debriefed and offered small compensation.

We crafted the content of the Tweets to be of relevance to the participants such that they meaningfully engage with the Tweet’s content (i.e., their responses are not arbitrary). We assumed participants understood the Amazon Alexa interfaces and were aware of the COVID-19 pandemic in general. However, we acknowledge that the level of interest regarding the COVID-19 vaccines could vary among the individual participants, affecting the extent to which their responses reflect their opinions. To assess the perceived accuracy, we used the questionnaire from \cite{Clayton}, adapted to the content presented in Figures 2-5. The questionnaire assesses the perceived accuracy of each of the Tweets on a 4-point Likert scale (not at all accurate, not very accurate, somewhat accurate, very accurate).  

To assess participants’ subjective attitudes and beliefs regarding the COVID-19 vaccine, we used a modified questionnaire from \cite{Biasio}. To assess the subjective attitudes we asked if the participants will receive a COVID-19 vaccine (Yes/No/I Don't Know); and (c) politically speaking, do they consider as (Republican/Democrat/Independent). The overall questionnaire was anonymous with no personally identifiable data collected from the participants. We utilized a 2x3 experimental design where participants will be randomized into one of six groups: (1) spoken back misleading Tweet with a follow up warning tag; (2) spoken back misleading Tweet without a warning tag suppressed by Alexa; (3) spoken back misleading Tweet preceded by an utterance of a warning cover; (4) spoken back verified Tweet; (5) spoken back verified Tweet with a follow up warning tag; and (6) verified Tweet preceded by an utterance of a warning cover.

\section{Results}
We conducted an online survey (N = 304) in January and February 2021. There were 195 (64.1\%) males and 103 (33.9\%) females, with 6 participants (2.0\%) identifying as non-cis. The age brackets in the sample were distributed as follows: 26 (8.6\%) [18 - 24], 112 (36.8\%) [25 - 34], 106 (34.9\%) [35 - 44], 35 (11.5\%) [45 - 54], 18 (5.9\%) [55 - 64], 6 (2.0\%) [65 - 74], and 1 (.3\%) [75 - 84].  Our sample was younger-leaning and slightly skewed towards male Twitter and Alexa users. The sample was also slightly Democrat-leaning with 51 (16.8\%) Republicans, 173 (56.9\%) Democrats, and 80 (26.3\%) Independent.

\subsection{Misperceptions: Accuracy of Tweets}
We first hypothesized that the utterance of a warning tag following a Tweet containing \textit{misleading} information (Figure 2a) about COVID-19 vaccines by Alexa will not reduce the perceived accuracy of the spoken back Tweet's content relative to a no warning tag condition (Figure 2b). We have to confirm this hypothesis since we haven't found any statistically significant result in the perceived accuracy ($U = 1288.5$, $p = .917, (\alpha = 0.05)$). In practical terms, this means that the follow up warning tag doesn't not work as intended - both groups perceived the Tweet as ``somewhat accurate'' on average. Alexa users either ignored the warning tag utterance at the end or perhaps the warning tag saying \texttt{Get the facts about COVID nineteen} is ambiguous and logically doesn't actually say that the particular Tweet is indeed misinformation. 

Granted, tagging the content is primarily intended for visual inspection and includes a warning exclamation favicon and a link to the COVID-19 verified information, which is not available directly to the users. We did allow for the participants to repeat the interaction with Alexa and change their reported perception of accuracy, but that didn't seem to matter. Perhaps, for a future use, Alexa could proceed and turn the warning tag into a question, prompting the user with \texttt{This Tweet was labeled as COVID nineteen misinformation. Would you like to hear the facts about the COVID nineteen?}. A case for such an adaptation also brings the test result for the third hypothesis, which was also didn't yield a significant result ($U = 1177$, $p = .490, (\alpha = 0.05)$). While it seems that the Alexa users correctly ignored the tag because the Twitter Reader skill inserted the warning tag utterance after it spoke back the Tweet as shown in Figure 4b, we don't know if this is because the Alexa users considered the warning tag at all. 

We also hypothesized that the utterance of a warning cover before a Tweet containing \textit{misleading} information about COVID-19 vaccines by Alexa (Figure 3) will not reduce the perceived accuracy of the spoken back Tweet's content relative to a no warning cover condition (Figure 3a). Here too, we didn't find a statistically significant result in the perceived accuracy ($U = 1280$, $p = .889, (\alpha = 0.05)$). This made us speculate if the perhaps the content of the Tweet was more so perceived as potentially an unverified rumor rather then a misinformation altogether (at least at the time of the study there was no official repudiation of the link between the COVID-19 vaccine and the death caused by the immune thrombocytopenia). What made us think of the nuanced choice of misinformation and COVID-19 rumors is the rejection of the fourth hypothesis where the utterance of the warning cover before Alexa spoke back the information from the valid Tweet shown on Figure 5 ($U = 1601$, $p = .002^{*}, (\alpha = 0.05)$, Cohen's $d =0.619$ medium). 

The cover was sufficiently potent for the users in the test group to perceive the following Tweet as ``not very accurate'' while the ones in the control condition as ''somewhat accurate'' as shown in Figure 6. This result suggest that the warning cover is a potent way of swaying users about potential misinformation, even if the following Tweet containing in fact verified information released by the CDC. This result, considering the possibility of developing a purely malicious Twitter Reader skill, could inject misinformation warning covers to a Tweet of choice and meddle with the public opinion on the COVID-19 vaccine effectiveness. A similar effort, though with actual Tweets and not warning labels, already surfaced on Twitter promoting a homegrown Russian vaccine and undercutting the vaccines from the rivals such as AstraZenecca \cite{Frenkel}.  

\begin{figure}[h]
  \centering
  \includegraphics[width=0.7\linewidth]{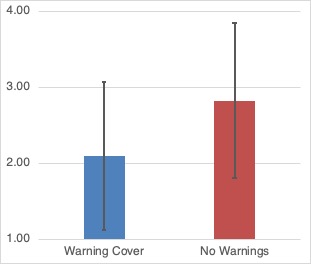}
  \caption{A difference in perceived accuracy between the group exposed to a warning cover utterance (Figure 5) and the group exposed only to the spoken back Tweet (Figure 4a)}
\end{figure}

\subsection{Misperceptions: Hesitancy}
Hesitancy of vaccination has been shown to be a discriminative lens through which people critically discern online information, which also includes COVID-19 Tweets on the COVID-19 topic \cite{Biasio}. Therefore, we wanted to test and see if the COVID-19 vaccine hesitancy will affect the perception of accuracy of a spoken back Tweet with \textit{misleading} information about COVID-19 in any condition (with a warning a preceding warning cover, with a following warning tag, or without any warning). When controlling for hesitancy, we found a significant difference only in the warning cover condition (Figure 3). The vaccine hesitant participants were more likely to perceive the spoken back misleading Tweet as ``somewhat accurate'', while the vaccine accepting and undecided as ``inaccurate'' as shown in Figure 7 ($\chi^{2}(2) = 10.058$, $p = .007^{*}, (\alpha = 0.05)$). It seems that the vaccine hesitant participants were alerted by the warning cover to contextually access the grim outlook linking the COVID-19 vaccine with death with their sceptic outlook of the COVID-19 vaccination, otherwise we would should have observed a similar result for the the warning tag condition. 

\begin{figure}[h]
  \centering
  \includegraphics[width=0.7\linewidth]{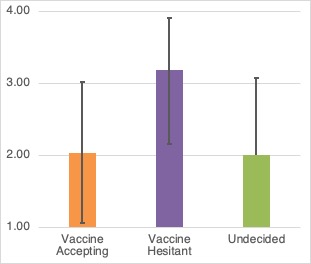}
  \caption{Vaccine hesitancy and perceived accuracy of the misleading Tweet in Figure 3.}
\end{figure}

That this is a plausible explanation of the obvious misperception suggest the similar test of the verified Tweet condition, where we also found a significant result only for the condition where a warning cover was uttered before the verified Twitter is spoken back by Alexa (Figure 5). Again, the vaccine hesitant participants were more likely to perceive the spoken back Tweet, this time containing verified information from the CDC, as ``somewhat accurate'', while the vaccine accepting and undecided as ``inaccurate'' as shown in Figure 8 ($\chi^{2}(2) = 6.432$, $p = .040^{*}, (\alpha = 0.05)$). We previously noted that the warning cover, spoken back by Alexa, seems sufficiently potent to add credibility to Alexa as the ``best friend forever'' \cite{Purington}. After all, the best friends don't lie and we trust them when they warn us that what they have found on Twitter might be a disinformation if they say so. But if this Twitter content aligns with our deeply held beliefs that the COVID-19 vaccines are bad and we for sure won't get vaccinated, we might disregard what Twitter has told Alexa to convey to us, and instead believe the content. After all, we might have heard that many users feel Twitter itself is biased and mislabeling content \cite {Geeng}, so who knows, maybe Alexa stumbled upon a mislabeled Tweet with an ominous COVID-19 misinformation warning cover. 

\begin{figure}[h]
  \centering
  \includegraphics[width=0.7\linewidth]{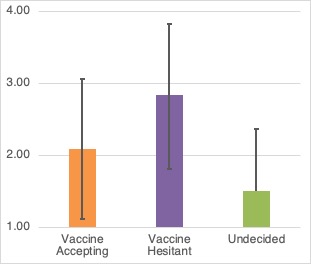}
  \caption{Vaccine hesitancy and perceived accuracy of the verified Tweet in Figure 5.}
\end{figure}

\subsection{Misperceptions: Political Leanings}
The COVID-19 pandemic was not ``immune'' from politicization, expectedly, as a highly polarizing topic. Curious to test the early evidence that perceptions are modulated by one's political leanings \cite{Zannettou, Christenson, Nyhan}, we controlled for participant's political identification to compare the perceived accuracy of the spoken back content by Alexa with and without warning covers/tags. We only found a statistical difference in for the condition where Alexa utters a warning cover before the valid Tweet (Figure 5). While the republican and democrat users took the Alexa ``warning'' and reported that this content is ``not so accurate'' the independent ones dismissed this warning and perceived this content as ''somewhat accurate'' as shown in Figure 9 ($\chi^{2}(2) = 6.171$, $p = .046^{*}, (\alpha = 0.05)$). 

\begin{figure}[h]
  \centering
  \includegraphics[width=0.7\linewidth]{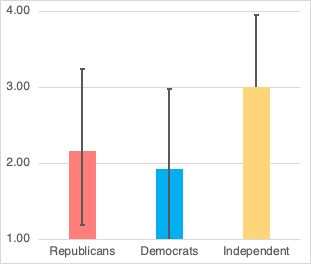}
  \caption{Vaccine hesitancy and perceived accuracy of the verified Tweet in Figure 5.}
\end{figure}

This result contrast the divided receptivity to allegedly misinformation content between the republicans (further drawn to believe it in presence of warnings) and democrats (repelled by the warnings) previously reported. But such a comparison might not hold relevant in our case first because the warnings and content in the previous studies focused on political social media content and second because the information interface was a screen, not a voice assistant. It might be expected for politically-laden users to assume the authority and trust Alexa as telling the truth so it comes to no surprise that they heeded the warning cover and payed little attention to the following Tweet content. The higher level of critical discernment for the independent might be the varied exposure to information to both left, center, and right leaning media and Twitter content in general \cite{Pennycook1}.  

\section{Discussion}
Exploring the soft moderation when communicated by Alexa instead of a label on a screen, we found that warning covers when converted from text to speech and uttered by Alexa work as intended in grabbing the attention of the Alexa users. All would have been good and promising, if it wasn't for the malicious skill inserting these covers as utterances before Alexa spoke back a Tweet content that was true and validated by the CDC. The users in our study trusted Alexa and it's seemingly benign third-party skill ``Twitter Reader'' to tell the truth without critically discerning the Tweets spoken back by Alexa. We explored what might cause such an effect and found a difference in perception based on one's political leanings - the users considering themselves independent were the ones actually trying to make sense of the Tweets regardless of what Alexa (e.g. Twitter) had to say about the content. Alexa in our study wasn't able to dispel any biases that were rooted in personal beliefs \cite{Mercandante2020}, \cite{Thorson}. One's personal hesitancy from COVID-19 vaccination sufficed for biased perception of the information from Alexa despite any labeling as long as the Tweets echoed their sceptic outlook of the whole COVID-19 vaccination business \cite{Pennycook, Nyhan}. 

\subsection{Ethical Implications}
While we set to investigate the effect of a third-party skill that arbitrary manipulates the soft moderation covers and tags attached by Twitter to dubious COVID-19 vaccine content and debriefed the participants at the end of the study, the results could have several ethical implications nonetheless. We exposed the participants to moderated content that could potentially affect participants' stance on COVID-19 vaccination and the pandemic. The exposure might not sway participants on their vaccine hesitancy in the long run, but could make the participants reconsider their approach of obtaining the vaccine for themselves or their families. The exposure to Twitter content spoken back by Alexa as a novel interface could also affect the participants' stance on communicating important topics with their voice assistants. A recent study found that personification of Alexa is associated with increased levels of satisfaction, so learning that Alexa could be secretly controlled by a malicious third-party skill to insert or drop Twitter misinformation warnings might affect the user satisfaction with Alexa \cite{Purington}. 

That a portion of the participants were able to critically discern the verified content even in presence of the maliciously inserted Alexa's warnings is reassuring and proves that users keep a critical mindset despite the proclivity for anthropomorphism towards Alexa \cite{Wagner}. However, the ease of crafting a malicious skill that could suppress or add warning utterances, could have unintended consequences. With the evidence of nation-states disseminating misinformation, it is possible that they could resort to malware for voice assistants to avoid both the soft and hard moderation on social media platforms like Twitter \cite{Thomas}. Sure, this could be far from the realm of possibility, even if the capabilities exist, but for such a sensitive topic as COVID-19 vaccination, meddling with spoken-back content from Alexa could give an edge to a vaccine competitor in the global race for development and procurement of a COVID-19 solution. We certainly condemn such ideas and such a misuse of our research results and therefore only presented a very limited proof-of-concept flow for such a malicious skill. For example, evidence for such a misinformation campaign has already surfaced on Twitter, promoting homegrown Russian vaccines and undercutting rivals \cite{Frenkel}. One could also point the malicious third-party skill to pull content from Parler instead of Twitter and package the disinformation vector as a skill that reads the ``real'' truth about the vaccine and avoid soft moderation altogether. Again, we condemn development of such a capability outside of academic research, even if it doesn't pose an imminent problem for the defense community \cite{Toby}.

\subsection{Future Research}
We acknowledge that there is further research to be done into investigating the manipulation of soft moderation and communication of Twitter content through voice assistants. Visually assessing warning labels might have a different effect when they are spoken by Alexa so it will be informative to see how a variation of the warning text could affect the perceived accuracy of content to which they are attached. Out results suggest that the follow up warning tag utterance has no effect at all due, we suspected, the ambiguous language. If perhaps Alexa speaks with a different tone and is more direct, saying \texttt{This Tweet seems like is spreading misinformation. Do you want me to check the COVID-19 facts on the official CDC website?}, one might hypothesize that the intended effect of soft moderation be better achieved. Certainly, the proposed adaptation entails extensive usable security research to determine the optimal way of delivering soft moderation warnings, especially with the option for Alexa to express emotions (e.g. ``disappointed'' and low volume in uttering the warning covers or ``disappointed'' and high volume in uttering the said tag \cite{AmazonEmo}). This is yet another step in future research, and we plan to broadly explore the domain of voice-based security warnings.

Along this line and the politicization of COVID-19 debate, an interesting experiment we plan to conduct is to go beyond the emotions of Alexa but instead test a voice application that reads the Tweets in the voice of a famous political person. Siri, Apple's voice assistant allows for customization with accents (e.g. Irish) so one would expect that the voice assistants in future could be configured to speak with the voice of our beloved celebrity or authoritative figures \cite{TTSAI}. A neural network that could synthesize speech directly from inputted text exists and with a moderate training models can be created to generate the speech from almost any individual \cite{NVIDIA-tacotron2}. It would be interesting to see how users will respond to same Tweets and warning covers read by John Oliver, Hillary Clinton, Donald Trump, or Barack Obama (one could already test these voices on the Vocodes website \cite{vocodes}).



\subsection{Combating Malicious Skills}
The misperception-inducing logic is enabled by customizing, in a relatively easy way, a blueprint template and registering a seemingly benign skill. As discussed in \cite{Malexa}, a thorough certification process could uncover the malicious logic and remove the skill from the Amazon Skills Store. Another solution is monitoring for suspicious skills' behaviour with a tool like the \textit{SkillExplorer} proposed in \cite{Guo}. Again, a skill can evade both certification and exploring by claiming that the rephrasing aims to communicate important COVID-19 vaccine information in an assistive way, for example, to non-native English speakers ~\cite{Jang}. 

As an additional layer of protection, feedback from users post-release could help close this gap. Twitter similarly hopes to identify and address misinformation through the use of pre-selected user ``fact checkers'', piloted in its Birdwatch program \cite{Ortutay2021}. Amazon could similarly crowd-source its Alexa skill moderation and allow users with a high ``helpfulness'' score, as in Birdwatch, to identify potentially malicious or misinforming skills for further review and removal. Though allowing users to flag skills may be helpful in eliminating misinformation, this crowd-sourced soft moderation could be exploited by malicious users to flag legitimate skills or hijacked by partisan users if a skill's content has been highly politicized.

In the context of malicious third party skills, instead of manipulating warning labels, a skill might be directed to an RSS feed that steadily promotes rumours and unverified COVID-19 vaccine information. One might not need to make a Twitter reader, but a Parler reader, to access a wealth of unverified claims about COVID-19 vaccination and supply Alexa with ``Parler COVID-19 briefings'' . For one, a widely shared information tidbit by ``influencers'' about the COVID-19 vaccine on Parler is that it contains HIV \cite{Aliapoulios}. We experimented with mostly rumours about COVID-19 in our study, but exposing participants to such blatant misinformation, spoken by a trusted intermediary (Alexa, and not Alex Jones), could possibly uncover important dynamics in the relationship or personification of Alexa as a ``Best Friend Forever'' \cite{Purington}.

\subsection{Scope Limitations}
The current study has important limitations. First, we limited our questions to couple of Tweets that were relevant to the state of the pandemic and mass immunization during the period of January-February 2021, which could be perceived with a different level of accuracy after a certain period of time. Overall, the findings may be specific to the effect the malicious skill has only on COVID-19 mass immunization and may not be generalizable to other topics. Second, though our survey asked participants whether they intend to receive a COVID-19 vaccine, we did not ask participants who answered in the affirmative how soon they intended to get vaccinated. An affirmative intention to vaccinate does not indicate an intention to vaccinate immediately and unconditionally, and therefore, the results cannot be interpreted as such. We likewise did not ask why participants who answered in the negative why they did not intend to get vaccinated or whether any factors could change this. A negative intention to vaccinate does not indicate an intention to never receive the COVID-19 vaccine, and likewise, these results should not be interpreted as such.

Third, regular Alexa or voice assistant participants in general may be desensitized to the spoken back information, which may have affected their perception of the COVID-19 vaccine irrespective of the rephrasing. Our experiment was limited to Alexa as a voice assistant of choice and the Twitter as a COVID-19 vaccination information. We were limited to evaluating the effects of COVID-19 vaccination only in the U.S., and this information might not be relevant for places where other vaccines (the AstraZeneca, Sinopharm or Gamaleya vaccines) are used. We were limited to the choice of soft moderation labels present at the time of the study that could change the perception of the spoken-back information if reworded later by Twitter. Finally, although we tried to sample a representative set of participants for our study using Amazon Mechanical Turk , the outcomes might have been different if we used another platform, or another type of sampling. Also, a larger sample size, one that was gender and politically balanced, could have provided a more nuanced view of Alexa as a ``Tweeter Reader'', but we had limited funding for this study.

\section{Conclusion}
In this study, we explored whether a third-party Alexa skill could successfully affect the perceived accuracy of COVID-19 vaccine Tweets and induce misperceptions in users by simply manipulating the soft moderation applied by Twitter. Additionally, we examined whether participants' vaccine hesitant sentiment and political leanings had any effect on the perception of accuracy of the Tweets spoken back by Alexa. Our findings suggest that users were most likely to be misled on Twitter information when Alexa utters a warning speech before it delivers the Tweet's content, regardless of it's validity. Participants' perception of the accuracy of both misleading and verified COVID-19 Tweets appear warped by participants' personal biases - participants judged Alexa's accuracy by how closely the response aligned with their own sceptic beliefs on the subject. We found a significant difference in perceived accuracy only for the apolitical Alexa users, which appear not to contextualize any misinformation labeling before they actually inspect a Tweet. All of our findings might be an effect of the lack of interaction or response users have when interacting with a voice assistant that they do not lack in online discourse on social media. Given the ease with which a user lacking developer experience can craft and share a third-party skill via the Amazon Skills Store, we believe it necessary to augment existing practices to catch malicious and misinforming skills like the one we showcased in this study. Likewise, we believe that the adaptation of the soft moderation with better verbal misinformation warnings may help break the confirmation bias feedback loop that reinforces listeners' biased vaccine outlooks.

\bibliographystyle{ACM-Reference-Format}
\bibliography{DLG-TwoTruthsAndALie}


\begin{thebibliography}{43}


\ifx \showCODEN    \undefined \def \showCODEN     #1{\unskip}     \fi
\ifx \showDOI      \undefined \def \showDOI       #1{#1}\fi
\ifx \showISBNx    \undefined \def \showISBNx     #1{\unskip}     \fi
\ifx \showISBNxiii \undefined \def \showISBNxiii  #1{\unskip}     \fi
\ifx \showISSN     \undefined \def \showISSN      #1{\unskip}     \fi
\ifx \showLCCN     \undefined \def \showLCCN      #1{\unskip}     \fi
\ifx \shownote     \undefined \def \shownote      #1{#1}          \fi
\ifx \showarticletitle \undefined \def \showarticletitle #1{#1}   \fi
\ifx \showURL      \undefined \def \showURL       {\relax}        \fi
\providecommand\bibfield[2]{#2}
\providecommand\bibinfo[2]{#2}
\providecommand\natexlab[1]{#1}
\providecommand\showeprint[2][]{arXiv:#2}

\bibitem[\protect\citeauthoryear{??}{NVI}{2021}]%
        {NVIDIA-tacotron2}
 \bibinfo{year}{2021}\natexlab{}.
\newblock
\newblock
\urldef\tempurl%
\url{https://github.com/NVIDIA/tacotron2}
\showURL{%
\tempurl}


\bibitem[\protect\citeauthoryear{??}{voi}{2021}]%
        {voiceflow}
 \bibinfo{year}{2021}\natexlab{}.
\newblock \bibinfo{title}{Visual Skill Building}.
\newblock
\newblock
\urldef\tempurl%
\url{https://www.voiceflow.com/}
\showURL{%
\tempurl}


\bibitem[\protect\citeauthoryear{??}{voc}{2021}]%
        {vocodes}
 \bibinfo{year}{2021}\natexlab{}.
\newblock \bibinfo{title}{Voice Encorder}.
\newblock
\newblock
\urldef\tempurl%
\url{https://vo.codes/#use}
\showURL{%
\tempurl}


\bibitem[\protect\citeauthoryear{Aliapoulios, Bevensee, Blackburn, Cristofaro,
  Stringhini, and Zannettou}{Aliapoulios et~al\mbox{.}}{2021}]%
        {Aliapoulios}
\bibfield{author}{\bibinfo{person}{Max Aliapoulios}, \bibinfo{person}{Emmi
  Bevensee}, \bibinfo{person}{Jeremy Blackburn}, \bibinfo{person}{Emiliano~De
  Cristofaro}, \bibinfo{person}{Gianluca Stringhini}, {and}
  \bibinfo{person}{Savvas Zannettou}.} \bibinfo{year}{2021}\natexlab{}.
\newblock \bibinfo{title}{An Early Look at the Parler Online Social Network}.
\newblock
\newblock
\showeprint[arxiv]{2101.03820v2}~[cs.SI]


\bibitem[\protect\citeauthoryear{Amazon}{Amazon}{2019}]%
        {alexa_skill_blueprints}
\bibfield{author}{\bibinfo{person}{Amazon}.} \bibinfo{year}{2019}\natexlab{}.
\newblock \bibinfo{title}{{Amazon Alexa | Skill Blueprints}}.
\newblock
\newblock
\newblock
\shownote{\url{https://blueprints.amazon.com/}.}


\bibitem[\protect\citeauthoryear{Biasio, Bonaccorsi, Lorini, and
  Pecorelli}{Biasio et~al\mbox{.}}{2020}]%
        {Biasio}
\bibfield{author}{\bibinfo{person}{Luigi~Roberto Biasio},
  \bibinfo{person}{Guglielmo Bonaccorsi}, \bibinfo{person}{Chiara Lorini},
  {and} \bibinfo{person}{Sergio Pecorelli}.} \bibinfo{year}{2020}\natexlab{}.
\newblock \showarticletitle{{Assessing COVID-19 vaccine literacy: A preliminary
  online survey}}.
\newblock \bibinfo{journal}{\emph{Human Vaccines \& Immunotherapeutics}}
  \bibinfo{volume}{0}, \bibinfo{number}{0} (\bibinfo{year}{2020}),
  \bibinfo{pages}{1--9}.
\newblock
\urldef\tempurl%
\url{https://doi.org/10.1080/21645515.2020.1829315}
\showDOI{\tempurl}


\bibitem[\protect\citeauthoryear{Boghardt}{Boghardt}{2009}]%
        {Boghardt}
\bibfield{author}{\bibinfo{person}{Thomas Boghardt}.}
  \bibinfo{year}{2009}\natexlab{}.
\newblock \bibinfo{title}{Operation INFEKTION: Soviet Bloc Intelligence and Its
  AIDS Disinformation Campaign}.
\newblock
\newblock
\urldef\tempurl%
\url{https://www.cia.gov/library/center-for-the-study-of-intelligence/csi-publications/csi-studies/studies/vol53no4/soviet-bloc-intelligence-and-its-aids.html}
\showURL{%
\tempurl}


\bibitem[\protect\citeauthoryear{Carlini, Mishra, Vaidya, Zhang, Sherr,
  Shields, Wagner, and Zhou}{Carlini et~al\mbox{.}}{2016}]%
        {Carlini}
\bibfield{author}{\bibinfo{person}{Nicholas Carlini}, \bibinfo{person}{Pratyush
  Mishra}, \bibinfo{person}{Tavish Vaidya}, \bibinfo{person}{Yuankai Zhang},
  \bibinfo{person}{Micah Sherr}, \bibinfo{person}{Clay Shields},
  \bibinfo{person}{David Wagner}, {and} \bibinfo{person}{Wenchao Zhou}.}
  \bibinfo{year}{2016}\natexlab{}.
\newblock \showarticletitle{Hidden Voice Commands}. In
  \bibinfo{booktitle}{\emph{25th {USENIX} Security Symposium ({USENIX} Security
  16)}}. \bibinfo{publisher}{{USENIX} Association}, \bibinfo{address}{Austin,
  TX}, \bibinfo{pages}{513--530}.
\newblock
\showISBNx{978-1-931971-32-4}
\urldef\tempurl%
\url{https://www.usenix.org/conference/usenixsecurity16/technical-sessions/presentation/carlini}
\showURL{%
\tempurl}


\bibitem[\protect\citeauthoryear{Chappell}{Chappell}{2021}]%
        {Chappell}
\bibfield{author}{\bibinfo{person}{Bill Chappell}.}
  \bibinfo{year}{2021}\natexlab{}.
\newblock \bibinfo{title}{{Instagram Bars Robert F. Kennedy Jr. For Spreading
  Vaccine Misinformation}}.
\newblock
\newblock
\newblock
\shownote{\url{https://www.npr.org/sections/coronavirus-live-updates/2021/02/11/966902737/instagram-bars-robert-f-kennedy-jr-for-spreading-vaccine-misinformation}.}


\bibitem[\protect\citeauthoryear{Christenson, Kreps, and Kriner}{Christenson
  et~al\mbox{.}}{2020}]%
        {Christenson}
\bibfield{author}{\bibinfo{person}{Dino~P Christenson},
  \bibinfo{person}{Sarah~E Kreps}, {and} \bibinfo{person}{Douglas~L Kriner}.}
  \bibinfo{year}{2020}\natexlab{}.
\newblock \showarticletitle{{Contemporary Presidency: Going Public in an Era of
  Social Media: Tweets, Corrections, and Public Opinion}}.
\newblock \bibinfo{journal}{\emph{Presidential Studies Quarterly}}
  (\bibinfo{year}{2020}).
\newblock


\bibitem[\protect\citeauthoryear{Clayton, Blair, Busam, Forstner, Glance,
  Green, Kawata, Kovvuri, Martin, Morgan, et~al\mbox{.}}{Clayton
  et~al\mbox{.}}{2019}]%
        {Clayton}
\bibfield{author}{\bibinfo{person}{Katherine Clayton}, \bibinfo{person}{Spencer
  Blair}, \bibinfo{person}{Jonathan~A Busam}, \bibinfo{person}{Samuel
  Forstner}, \bibinfo{person}{John Glance}, \bibinfo{person}{Guy Green},
  \bibinfo{person}{Anna Kawata}, \bibinfo{person}{Akhila Kovvuri},
  \bibinfo{person}{Jonathan Martin}, \bibinfo{person}{Evan Morgan},
  {et~al\mbox{.}}} \bibinfo{year}{2019}\natexlab{}.
\newblock \showarticletitle{Real solutions for fake news? Measuring the
  effectiveness of general warnings and fact-check tags in reducing belief in
  false stories on social media}.
\newblock \bibinfo{journal}{\emph{Political Behavior}} (\bibinfo{year}{2019}),
  \bibinfo{pages}{1--23}.
\newblock


\bibitem[\protect\citeauthoryear{Featherstone, Barnett, Ruiz, Zhuang, and
  Millam}{Featherstone et~al\mbox{.}}{2020}]%
        {Featherstone20}
\bibfield{author}{\bibinfo{person}{Jieyu~D. Featherstone},
  \bibinfo{person}{George~A. Barnett}, \bibinfo{person}{Jeanette~B. Ruiz},
  \bibinfo{person}{Yurong Zhuang}, {and} \bibinfo{person}{Benjamin~J. Millam}.}
  \bibinfo{year}{2020}\natexlab{}.
\newblock \showarticletitle{Exploring childhood anti-vaccine and pro-vaccine
  communities on twitter – a perspective from influential users}.
\newblock \bibinfo{journal}{\emph{Online Social Networks and Media}}
  \bibinfo{volume}{20} (\bibinfo{year}{2020}), \bibinfo{pages}{100105}.
\newblock
\showISSN{2468-6964}
\urldef\tempurl%
\url{https://doi.org/10.1016/j.osnem.2020.100105}
\showDOI{\tempurl}


\bibitem[\protect\citeauthoryear{for Disease~Control and Prevention}{for
  Disease~Control and Prevention}{2021}]%
        {CDC}
\bibfield{author}{\bibinfo{person}{Centers for Disease~Control} {and}
  \bibinfo{person}{Prevention}.} \bibinfo{year}{2021}\natexlab{}.
\newblock \bibinfo{title}{{COVID-19 Vaccines and Allergic Reactions}}.
\newblock
\newblock
\newblock
\shownote{\url{https://www.cdc.gov/coronavirus/2019-ncov/vaccines/safety/allergic-reaction.html}.}


\bibitem[\protect\citeauthoryear{Frenkel, Abi-Habib, and Barnes}{Frenkel
  et~al\mbox{.}}{2021}]%
        {Frenkel}
\bibfield{author}{\bibinfo{person}{Sheera Frenkel}, \bibinfo{person}{Maria
  Abi-Habib}, {and} \bibinfo{person}{Julian~E. Barnes}.}
  \bibinfo{year}{2021}\natexlab{}.
\newblock \bibinfo{title}{{Russian Campaign Promotes Homegrown Vaccine and
  Undercuts Rivals}}.
\newblock
\newblock
\newblock
\shownote{\url{https://www.nytimes.com/2021/02/05/technology/russia-covid-vaccine-disinformation.html}.}


\bibitem[\protect\citeauthoryear{{Gao, Catherine}}{{Gao, Catherine}}{2019}]%
        {AmazonEmo}
\bibfield{author}{\bibinfo{person}{{Gao, Catherine}}.}
  \bibinfo{year}{2019}\natexlab{}.
\newblock \bibinfo{title}{{Use New Alexa Emotions and Speaking Styles to Create
  a More Natural and Intuitive Voice Experience}}.
\newblock
\newblock
\newblock
\shownote{\url{https://developer.amazon.com/en-US/blogs/alexa/alexa-skills-kit/2019/11/new-alexa-emotions-and-speaking-styles}.}


\bibitem[\protect\citeauthoryear{Geeng, Francisco, West, and Roesner}{Geeng
  et~al\mbox{.}}{2020}]%
        {Geeng}
\bibfield{author}{\bibinfo{person}{Christine Geeng}, \bibinfo{person}{Tiona
  Francisco}, \bibinfo{person}{Jevin West}, {and} \bibinfo{person}{Franziska
  Roesner}.} \bibinfo{year}{2020}\natexlab{}.
\newblock \bibinfo{title}{Social Media COVID-19 Misinformation Interventions
  Viewed Positively, But Have Limited Impact}.
\newblock
\newblock
\showeprint[arxiv]{2012.11055}~[cs.CY]


\bibitem[\protect\citeauthoryear{Guo, Lin, Li, and Chen}{Guo
  et~al\mbox{.}}{2020}]%
        {Guo}
\bibfield{author}{\bibinfo{person}{Zhixiu Guo}, \bibinfo{person}{Zijin Lin},
  \bibinfo{person}{Pan Li}, {and} \bibinfo{person}{Kai Chen}.}
  \bibinfo{year}{2020}\natexlab{}.
\newblock \showarticletitle{SkillExplorer: Understanding the Behavior of Skills
  in Large Scale}. In \bibinfo{booktitle}{\emph{29th {USENIX} Security
  Symposium ({USENIX} Security 20)}}. \bibinfo{publisher}{{USENIX}
  Association}, \bibinfo{pages}{2649--2666}.
\newblock
\showISBNx{978-1-939133-17-5}
\urldef\tempurl%
\url{https://www.usenix.org/conference/usenixsecurity20/presentation/guo}
\showURL{%
\tempurl}


\bibitem[\protect\citeauthoryear{Jachim, Sharevski, and Treebridge}{Jachim
  et~al\mbox{.}}{2020}]%
        {Jachim}
\bibfield{author}{\bibinfo{person}{Peter Jachim}, \bibinfo{person}{Filipo
  Sharevski}, {and} \bibinfo{person}{Paige Treebridge}.}
  \bibinfo{year}{2020}\natexlab{}.
\newblock \showarticletitle{TrollHunter [Evader]: Automated Detection [Evasion]
  of Twitter Trolls During the COVID-19 Pandemic}. In
  \bibinfo{booktitle}{\emph{New Security Paradigms Workshop 2020}} (Online,
  USA) \emph{(\bibinfo{series}{NSPW '20})}. \bibinfo{publisher}{Association for
  Computing Machinery}, \bibinfo{address}{New York, NY, USA},
  \bibinfo{pages}{59–75}.
\newblock
\showISBNx{9781450389952}
\urldef\tempurl%
\url{https://doi.org/10.1145/3442167.3442169}
\showDOI{\tempurl}


\bibitem[\protect\citeauthoryear{Jang, Song, Chung, Wang, and Lee}{Jang
  et~al\mbox{.}}{2014}]%
        {Jang}
\bibfield{author}{\bibinfo{person}{Yeongjin Jang}, \bibinfo{person}{Chengyu
  Song}, \bibinfo{person}{Simon~P. Chung}, \bibinfo{person}{Tielei Wang}, {and}
  \bibinfo{person}{Wenke Lee}.} \bibinfo{year}{2014}\natexlab{}.
\newblock \showarticletitle{{A11Y Attacks: Exploiting Accessibility in
  Operating Systems}}. In \bibinfo{booktitle}{\emph{Proceedings of the 2014 ACM
  SIGSAC Conference on Computer and Communications Security}} (Scottsdale,
  Arizona, USA) \emph{(\bibinfo{series}{CCS '14})}. \bibinfo{publisher}{ACM},
  \bibinfo{address}{New York, NY, USA}, \bibinfo{pages}{103--115}.
\newblock
\showISBNx{978-1-4503-2957-6}
\urldef\tempurl%
\url{https://doi.org/10.1145/2660267.2660295}
\showDOI{\tempurl}


\bibitem[\protect\citeauthoryear{Kumar, Paccagnella, Murley, Hennenfent, Mason,
  Bates, and Bailey}{Kumar et~al\mbox{.}}{2018}]%
        {Kumar}
\bibfield{author}{\bibinfo{person}{Deepak Kumar}, \bibinfo{person}{Riccardo
  Paccagnella}, \bibinfo{person}{Paul Murley}, \bibinfo{person}{Eric
  Hennenfent}, \bibinfo{person}{Joshua Mason}, \bibinfo{person}{Adam Bates},
  {and} \bibinfo{person}{Michael Bailey}.} \bibinfo{year}{2018}\natexlab{}.
\newblock \showarticletitle{Skill Squatting Attacks on Amazon Alexa}. In
  \bibinfo{booktitle}{\emph{27th {USENIX} Security Symposium ({USENIX} Security
  18)}}. \bibinfo{publisher}{{USENIX} Association},
  \bibinfo{address}{Baltimore, MD}, \bibinfo{pages}{33--47}.
\newblock
\showISBNx{978-1-939133-04-5}
\urldef\tempurl%
\url{https://www.usenix.org/conference/usenixsecurity18/presentation/kumar}
\showURL{%
\tempurl}


\bibitem[\protect\citeauthoryear{Lau, Zimmerman, and Schaub}{Lau
  et~al\mbox{.}}{2018}]%
        {Lau}
\bibfield{author}{\bibinfo{person}{Josephine Lau}, \bibinfo{person}{Benjamin
  Zimmerman}, {and} \bibinfo{person}{Florian Schaub}.}
  \bibinfo{year}{2018}\natexlab{}.
\newblock \showarticletitle{{Alexa, Are You Listening?: Privacy Perceptions,
  Concerns and Privacy-seeking Behaviors with Smart Speakers}}.
\newblock \bibinfo{journal}{\emph{Proc. ACM Hum.-Comput. Interact.}}
  \bibinfo{volume}{2}, \bibinfo{number}{CSCW}, Article \bibinfo{articleno}{102}
  (\bibinfo{date}{Nov.} \bibinfo{year}{2018}), \bibinfo{numpages}{31}~pages.
\newblock
\showISSN{2573-0142}
\urldef\tempurl%
\url{https://doi.org/10.1145/3274371}
\showDOI{\tempurl}


\bibitem[\protect\citeauthoryear{Mercadante and Law}{Mercadante and
  Law}{2020}]%
        {Mercandante2020}
\bibfield{author}{\bibinfo{person}{Amanda~R. Mercadante} {and}
  \bibinfo{person}{Anandi~V. Law}.} \bibinfo{year}{2020}\natexlab{}.
\newblock \showarticletitle{Will they, or Won't they? Examining patients'
  vaccine intention for flu and COVID-19 using the Health Belief Model}.
\newblock \bibinfo{journal}{\emph{Research in Social and Administrative
  Pharmacy}} (\bibinfo{year}{2020}).
\newblock
\showISSN{1551-7411}
\urldef\tempurl%
\url{https://doi.org/10.1016/j.sapharm.2020.12.012}
\showDOI{\tempurl}


\bibitem[\protect\citeauthoryear{Nyhan and Reifler}{Nyhan and Reifler}{2010}]%
        {Nyhan}
\bibfield{author}{\bibinfo{person}{Brendan Nyhan} {and} \bibinfo{person}{Jason
  Reifler}.} \bibinfo{year}{2010}\natexlab{}.
\newblock \showarticletitle{When corrections fail: The persistence of political
  misperceptions}.
\newblock \bibinfo{journal}{\emph{Political Behavior}} \bibinfo{volume}{32},
  \bibinfo{number}{2} (\bibinfo{year}{2010}), \bibinfo{pages}{303--330}.
\newblock


\bibitem[\protect\citeauthoryear{Ortutay}{Ortutay}{2021}]%
        {Ortutay2021}
\bibfield{author}{\bibinfo{person}{Barbara Ortutay}.}
  \bibinfo{year}{2021}\natexlab{}.
\newblock \showarticletitle{Twitter launches crowd-sourced fact-checking
  project}.
\newblock \bibinfo{journal}{\emph{Associated Press - AP News}}
  (\bibinfo{year}{2021}).
\newblock
\urldef\tempurl%
\url{https://apnews.com/article/twitter-launch-crowd-sourced-fact-check-589809d4c9a7eceda1ea8293b0a14af2}
\showURL{%
\tempurl}


\bibitem[\protect\citeauthoryear{Peironi, Jachim, Jachim, and
  Sharevski}{Peironi et~al\mbox{.}}{2021}]%
        {Pieroni}
\bibfield{author}{\bibinfo{person}{Emma Peironi}, \bibinfo{person}{Peter
  Jachim}, \bibinfo{person}{Nathaniel Jachim}, {and} \bibinfo{person}{Filipo
  Sharevski}.} \bibinfo{year}{2021}\natexlab{}.
\newblock \showarticletitle{Parlermonium: A Data-Driven UX Design Evaluation of
  the Parler Platform}. In \bibinfo{booktitle}{\emph{Critical Thinking in the
  Age of Misinformation CHI 2021}}.
\newblock


\bibitem[\protect\citeauthoryear{Pennycook, Bear, Collins, and Rand}{Pennycook
  et~al\mbox{.}}{2020}]%
        {Pennycook}
\bibfield{author}{\bibinfo{person}{Gordon Pennycook}, \bibinfo{person}{Adam
  Bear}, \bibinfo{person}{Evan~T Collins}, {and} \bibinfo{person}{David~G
  Rand}.} \bibinfo{year}{2020}\natexlab{}.
\newblock \showarticletitle{The implied truth effect: Attaching warnings to a
  subset of fake news headlines increases perceived accuracy of headlines
  without warnings}.
\newblock \bibinfo{journal}{\emph{Management Science}} (\bibinfo{year}{2020}).
\newblock


\bibitem[\protect\citeauthoryear{Pennycook, Cannon, and Rand}{Pennycook
  et~al\mbox{.}}{2018}]%
        {Pennycook1}
\bibfield{author}{\bibinfo{person}{Gordon Pennycook}, \bibinfo{person}{Tyrone~D
  Cannon}, {and} \bibinfo{person}{David~G Rand}.}
  \bibinfo{year}{2018}\natexlab{}.
\newblock \showarticletitle{Prior exposure increases perceived accuracy of fake
  news.}
\newblock \bibinfo{journal}{\emph{Journal of experimental psychology: general}}
  \bibinfo{volume}{147}, \bibinfo{number}{12} (\bibinfo{year}{2018}),
  \bibinfo{pages}{1865}.
\newblock


\bibitem[\protect\citeauthoryear{Purington, Taft, Sannon, Bazarova, and
  Taylor}{Purington et~al\mbox{.}}{2017}]%
        {Purington}
\bibfield{author}{\bibinfo{person}{Amanda Purington},
  \bibinfo{person}{Jessie~G. Taft}, \bibinfo{person}{Shruti Sannon},
  \bibinfo{person}{Natalya~N. Bazarova}, {and} \bibinfo{person}{Samuel~Hardman
  Taylor}.} \bibinfo{year}{2017}\natexlab{}.
\newblock \showarticletitle{"Alexa is My New BFF": Social Roles, User
  Satisfaction, and Personification of the Amazon Echo}. In
  \bibinfo{booktitle}{\emph{CHI Extended Abstracts}} (Denver, Colorado, USA)
  \emph{(\bibinfo{series}{CHI EA '17})}. \bibinfo{publisher}{Association for
  Computing Machinery}, \bibinfo{address}{New York, NY, USA},
  \bibinfo{pages}{2853–2859}.
\newblock
\showISBNx{9781450346566}
\urldef\tempurl%
\url{https://doi.org/10.1145/3027063.3053246}
\showDOI{\tempurl}


\bibitem[\protect\citeauthoryear{Roth and Pickles}{Roth and Pickles}{2020}]%
        {Roth}
\bibfield{author}{\bibinfo{person}{Yoel Roth} {and} \bibinfo{person}{Nick
  Pickles}.} \bibinfo{year}{2020}\natexlab{}.
\newblock \bibinfo{title}{Updating our approach to misleading information}.
\newblock
\newblock
\urldef\tempurl%
\url{https://blog.twitter.com/en\_us/topics/product/2020/updating-our-approach-to-misleading-information.html}
\showURL{%
\tempurl}


\bibitem[\protect\citeauthoryear{{Security Research Labs}}{{Security Research
  Labs}}{2019}]%
        {srl}
\bibfield{author}{\bibinfo{person}{{Security Research Labs}}.}
  \bibinfo{year}{2019}\natexlab{}.
\newblock \bibinfo{title}{{Smart Spies: Alexa and Google Home expose users to
  vishing and eavesdropping}}.
\newblock \bibinfo{howpublished}{\url{https://srlabs.de/bites/smart-spies/}}.
\newblock


\bibitem[\protect\citeauthoryear{Sharevski, Jachim, and Florek}{Sharevski
  et~al\mbox{.}}{2020a}]%
        {mimTweet}
\bibfield{author}{\bibinfo{person}{Filipo Sharevski}, \bibinfo{person}{Peter
  Jachim}, {and} \bibinfo{person}{Kevin Florek}.}
  \bibinfo{year}{2020}\natexlab{a}.
\newblock \showarticletitle{To Tweet or Not to Tweet: Covertly Manipulating a
  Twitter Debate on Vaccines Using Malware-Induced Misperceptions}. In
  \bibinfo{booktitle}{\emph{Proceedings of the 15th International Conference on
  Availability, Reliability and Security}} (Virtual Event, Ireland)
  \emph{(\bibinfo{series}{ARES '20})}. \bibinfo{publisher}{Association for
  Computing Machinery}, \bibinfo{address}{New York, NY, USA}, Article
  \bibinfo{articleno}{75}, \bibinfo{numpages}{12}~pages.
\newblock
\showISBNx{9781450388337}
\urldef\tempurl%
\url{https://doi.org/10.1145/3407023.3407025}
\showDOI{\tempurl}


\bibitem[\protect\citeauthoryear{Sharevski, Jachim, Treebridge, Li, Babin, and
  Adadevoh}{Sharevski et~al\mbox{.}}{2021}]%
        {Malexa}
\bibfield{author}{\bibinfo{person}{Filipo Sharevski}, \bibinfo{person}{Peter
  Jachim}, \bibinfo{person}{Paige Treebridge}, \bibinfo{person}{Audrey Li},
  \bibinfo{person}{Adam Babin}, {and} \bibinfo{person}{Christopher Adadevoh}.}
  \bibinfo{year}{2021}\natexlab{}.
\newblock \showarticletitle{Meet Malexa, Alexa's malicious twin:
  Malware-induced misperception through intelligent voice assistants}.
\newblock \bibinfo{journal}{\emph{International Journal of Human-Computer
  Studies}}  \bibinfo{volume}{149} (\bibinfo{year}{2021}),
  \bibinfo{pages}{102604}.
\newblock
\showISSN{1071-5819}
\urldef\tempurl%
\url{https://doi.org/10.1016/j.ijhcs.2021.102604}
\showDOI{\tempurl}


\bibitem[\protect\citeauthoryear{Sharevski, Treebridge, Jachim, Li, Babin, and
  Westbrook}{Sharevski et~al\mbox{.}}{2020b}]%
        {mimfacebook}
\bibfield{author}{\bibinfo{person}{Filipo Sharevski}, \bibinfo{person}{Paige
  Treebridge}, \bibinfo{person}{Peter Jachim}, \bibinfo{person}{Audrey Li},
  \bibinfo{person}{Adam Babin}, {and} \bibinfo{person}{Jessica Westbrook}.}
  \bibinfo{year}{2020}\natexlab{b}.
\newblock \bibinfo{title}{Beyond Trolling: Malware-Induced Misperception
  Attacks on Polarized Facebook Discourse}.
\newblock
\newblock
\showeprint[arxiv]{2002.03885}~[cs.HC]


\bibitem[\protect\citeauthoryear{Sharevski, Treebridge, Jachim, Li, Babin, and
  Westbrook}{Sharevski et~al\mbox{.}}{2020c}]%
        {Sharevski}
\bibfield{author}{\bibinfo{person}{Filipo Sharevski}, \bibinfo{person}{Paige
  Treebridge}, \bibinfo{person}{Peter Jachim}, \bibinfo{person}{Audrey Li},
  \bibinfo{person}{Adam Babin}, {and} \bibinfo{person}{Jessica Westbrook}.}
  \bibinfo{year}{2020}\natexlab{c}.
\newblock \bibinfo{title}{Meet Malexa, Alexa's Malicious Twin: Malware-Induced
  Misperception Through Intelligent Voice Assistants}.
\newblock
\newblock
\showeprint[arxiv]{2002.03466}~[cs.CR]


\bibitem[\protect\citeauthoryear{{Shen}, {Pang}, {Weiss}, {Schuster}, {Jaitly},
  {Yang}, {Chen}, {Zhang}, {Wang}, {Skerrv-Ryan}, {Saurous},
  {Agiomvrgiannakis}, and {Wu}}{{Shen} et~al\mbox{.}}{2018}]%
        {TTSAI}
\bibfield{author}{\bibinfo{person}{J. {Shen}}, \bibinfo{person}{R. {Pang}},
  \bibinfo{person}{R.~J. {Weiss}}, \bibinfo{person}{M. {Schuster}},
  \bibinfo{person}{N. {Jaitly}}, \bibinfo{person}{Z. {Yang}},
  \bibinfo{person}{Z. {Chen}}, \bibinfo{person}{Y. {Zhang}},
  \bibinfo{person}{Y. {Wang}}, \bibinfo{person}{R. {Skerrv-Ryan}},
  \bibinfo{person}{R.~A. {Saurous}}, \bibinfo{person}{Y. {Agiomvrgiannakis}},
  {and} \bibinfo{person}{Y. {Wu}}.} \bibinfo{year}{2018}\natexlab{}.
\newblock \showarticletitle{Natural TTS Synthesis by Conditioning Wavenet on
  MEL Spectrogram Predictions}. In \bibinfo{booktitle}{\emph{2018 IEEE
  International Conference on Acoustics, Speech and Signal Processing
  (ICASSP)}}. \bibinfo{pages}{4779--4783}.
\newblock
\urldef\tempurl%
\url{https://doi.org/10.1109/ICASSP.2018.8461368}
\showDOI{\tempurl}


\bibitem[\protect\citeauthoryear{Shevlane and Dafoe}{Shevlane and
  Dafoe}{2020}]%
        {Toby}
\bibfield{author}{\bibinfo{person}{Toby Shevlane} {and} \bibinfo{person}{Allan
  Dafoe}.} \bibinfo{year}{2020}\natexlab{}.
\newblock \showarticletitle{The Offense-Defense Balance of Scientific
  Knowledge: Does Publishing AI Research Reduce Misuse?}. In
  \bibinfo{booktitle}{\emph{Proceedings of the AAAI/ACM Conference on AI,
  Ethics, and Society}} (New York, NY, USA) \emph{(\bibinfo{series}{AIES
  '20})}. \bibinfo{publisher}{Association for Computing Machinery},
  \bibinfo{address}{New York, NY, USA}, \bibinfo{pages}{173–179}.
\newblock
\showISBNx{9781450371100}
\urldef\tempurl%
\url{https://doi.org/10.1145/3375627.3375815}
\showDOI{\tempurl}


\bibitem[\protect\citeauthoryear{Starbird}{Starbird}{2017}]%
        {Starbird}
\bibfield{author}{\bibinfo{person}{Kate Starbird}.}
  \bibinfo{year}{2017}\natexlab{}.
\newblock \showarticletitle{Examining the alternative media ecosystem through
  the production of alternative narratives of mass shooting events on Twitter.}
\newblock


\bibitem[\protect\citeauthoryear{Thomas, Grier, and Paxson}{Thomas
  et~al\mbox{.}}{2012}]%
        {Thomas}
\bibfield{author}{\bibinfo{person}{Kurt Thomas}, \bibinfo{person}{Chris Grier},
  {and} \bibinfo{person}{Vern Paxson}.} \bibinfo{year}{2012}\natexlab{}.
\newblock \showarticletitle{Adapting social spam infrastructure for political
  censorship}. In \bibinfo{booktitle}{\emph{5th $\{$USENIX$\}$ Workshop on
  Large-Scale Exploits and Emergent Threats ($\{$LEET$\}$ 12)}}.
\newblock


\bibitem[\protect\citeauthoryear{Thorson}{Thorson}{2016}]%
        {Thorson}
\bibfield{author}{\bibinfo{person}{Emily Thorson}.}
  \bibinfo{year}{2016}\natexlab{}.
\newblock \showarticletitle{Belief echoes: The persistent effects of corrected
  misinformation}.
\newblock \bibinfo{journal}{\emph{Political Communication}}
  \bibinfo{volume}{33}, \bibinfo{number}{3} (\bibinfo{year}{2016}),
  \bibinfo{pages}{460--480}.
\newblock


\bibitem[\protect\citeauthoryear{Vaidya, Zhang, Sherr, and Shields}{Vaidya
  et~al\mbox{.}}{2015}]%
        {Vaidya}
\bibfield{author}{\bibinfo{person}{Tavish Vaidya}, \bibinfo{person}{Yuankai
  Zhang}, \bibinfo{person}{Micah Sherr}, {and} \bibinfo{person}{Clay Shields}.}
  \bibinfo{year}{2015}\natexlab{}.
\newblock \showarticletitle{Cocaine Noodles: Exploiting the Gap between Human
  and Machine Speech Recognition}. In \bibinfo{booktitle}{\emph{9th {USENIX}
  Workshop on Offensive Technologies ({WOOT} 15)}}.
  \bibinfo{publisher}{{USENIX} Association}, \bibinfo{address}{Washington,
  D.C.}
\newblock
\urldef\tempurl%
\url{https://www.usenix.org/conference/woot15/workshop-program/presentation/vaidya}
\showURL{%
\tempurl}


\bibitem[\protect\citeauthoryear{Wagner and Schramm-Klein}{Wagner and
  Schramm-Klein}{2019}]%
        {Wagner}
\bibfield{author}{\bibinfo{person}{Katja Wagner} {and} \bibinfo{person}{Hanna
  Schramm-Klein}.} \bibinfo{year}{2019}\natexlab{}.
\newblock \showarticletitle{Alexa, are you human? Investigating
  anthropomorphism of digital voice assistants--A qualitative approach}.
\newblock \bibinfo{journal}{\emph{Robot Interactions and Interfaces}}
  (\bibinfo{year}{2019}).
\newblock


\bibitem[\protect\citeauthoryear{Yuan, Chen, Zhao, Long, Liu, Chen, Zhang,
  Huang, Wang, and Gunter}{Yuan et~al\mbox{.}}{2018}]%
        {Xuejing}
\bibfield{author}{\bibinfo{person}{Xuejing Yuan}, \bibinfo{person}{Yuxuan
  Chen}, \bibinfo{person}{Yue Zhao}, \bibinfo{person}{Yunhui Long},
  \bibinfo{person}{Xiaokang Liu}, \bibinfo{person}{Kai Chen},
  \bibinfo{person}{Shengzhi Zhang}, \bibinfo{person}{Heqing Huang},
  \bibinfo{person}{XiaoFeng Wang}, {and} \bibinfo{person}{Carl~A. Gunter}.}
  \bibinfo{year}{2018}\natexlab{}.
\newblock \showarticletitle{Commandersong: A Systematic Approach for Practical
  Adversarial Voice Recognition}. In \bibinfo{booktitle}{\emph{Proceedings of
  the 27th USENIX Conference on Security Symposium}} (Baltimore, MD, USA)
  \emph{(\bibinfo{series}{SEC'18})}. \bibinfo{publisher}{USENIX Association},
  \bibinfo{address}{USA}, \bibinfo{pages}{49--64}.
\newblock
\showISBNx{9781931971461}


\bibitem[\protect\citeauthoryear{Zannettou}{Zannettou}{2021}]%
        {Zannettou}
\bibfield{author}{\bibinfo{person}{Savvas Zannettou}.}
  \bibinfo{year}{2021}\natexlab{}.
\newblock \showarticletitle{{``I Won the Election!'':An Empirical Analysis of
  Soft Moderation Interventions on Twitter}}.
\newblock \bibinfo{journal}{\emph{arXiv}}  \bibinfo{volume}{2101.07183v1}
  (\bibinfo{date}{18 January} \bibinfo{year}{2021}).
\newblock
\newblock
\shownote{\url{https://arxiv.org/pdf/2101.07183.pdf}.}


\end{thebibliography}

\end{document}